# Zeros in Bosonic Wave-Function Result in Local Anti-Bunching: Refining Feynman's argument

Avi Marchewka and Er'el Granot

*Department of Electrical and Electronics Engineering, Ariel University of Samaria, Ariel, Israel*

**Abstract**

The effect of boson bunching is frequently mentioned and discussed in the literature. This effect is the manifestation of bosons tendency to "travel" in clusters. One of the core arguments for boson bunching was formulated by Feynman in his well-known lecture series and has been frequently used ever since. By comparing the scattering probabilities of two bosons and of two non-identical particles, Feynman concluded: "We have the result that it is *twice as likely* to find two *identical* Bose particles scattered into the same state *as you would calculate assuming the particles were different*." [1]

Indeed, in most scenarios, this reasoning is valid, however, as it is shown in this paper, there are cases, even in the most ordinary scattering scenarios, where this reasoning is invalid, and in fact the opposite occurs: boson anti-bunching appears. Similarly, it is shown that at exactly the same scenarios – fermions bunch together.

PACS indexes: 05.30.Jp, 03.65.Ta, 03.65.Nk, 05.30.Fk

**Introduction**

It is well-known that the wave function of non-interacting particles consists of single particle wave-functions (SPWF). If the particles are non-identical then the joint wave-function is a product of SPWF's.

However, if the particles are identical then symmetrization under particle exchange is required. The joint wave-function should be symmetric for boson particles and anti-symmetric for fermion particles[2,3].

It is of fundamental importance that the overall symmetry carries some intrinsic property, which is independent of the specific SPWF.

Boson bunching as well as fermion anti-bunching were regarded as such properties e.g. [1-2,4], (for the sake of simplicity, we will follow Feynman's description, which uses the terms:"fermions" and "bosons", when it should have been "spinless fermions" and "spinless bosons" respectively.), since its explanation in the literature was totally independent of the specific SPWF. Hence, bunching was regarded as one of bosons' properties.

However, in this paper we reexamine the explanation commonly used in the literature and show that bosons do not always bunch, in fact, in some cases they even anti-bunch. Therefore, bunching is not an intrinsic property that is introduced by the symmetry of the wave function, and is dependent of the specific SPWF.

**Bosons bunching**

In 1961 U. Fano gave a simple quantum explanation[5,6, and 7]) to the 1956 Hanbury Brown and Twiss (HBT) photon-bunching effect[8].

His explanation consisted of two (far) point sources and two (close) detectors. It was shown that the bunching probability is twice what would be expected from classical particles.

Almost during the same time, R. Feynman in his 1961-3 lectures on physics[1,4] presented a very similar explanation to boson bunching, which was based on a very similar model consisting of two separated sources and two closed detectors. Similar models, with related diagrams, have also been used in numerous places: in elementary particles physics [1,9], atomic physics [4,10], and optics [11,12].

Feynman compared the dynamics of non-identical particles to identical bosons. Let's follow his logic.

In the two cases (non-identical particles and identical bosons[13]) two particles are emitted from two distinct sources $\alpha, \beta$ and detected on two different but closely placed detectors $1, 2$ (Fig.1).

Following Feynman's terminology, $a_1$ and $a_2$ are the amplitudes for particle $a$ to scatter to detector 1 and 2 respectively, while $b_1$ and $b_2$ are the corresponding amplitudes for particle b to reach detector 1 and 2 respectively.

If the particles are not identical then the amplitude for both particles to be detected is either $a_1 b_2$ or $a_2 b_1$, and therefore, due to their distinguishability, the probability for this event is

$$|a_1 b_2|^2 + |a_2 b_1|^2. \tag{1}$$

Now, if we suppose that the two detectors 1 and 2 are very close together then

$a_1 \cong a_2 \cong a$, and $b_1 \cong b_2 \cong b$, hence the probability density is approximately

$$p \cong 2|ab|^2. \tag{2}$$

However, if the particles are identical bosons, then the two amplitudes should be added $a_1 b_2 + a_2 b_1$ and the probability density is

$$p = |a_1 b_2 + a_2 b_1|^2. \tag{3}$$

Now, again, if the two detectors are very close together then the probability density can be approximated to

$$p \cong 4|ab|^2. \tag{4}$$

Feynman therefore concluded that it is twice as likely to find two identical bosons scattered in the same location as one would calculate assuming the particles were different. In his lecture, as well as in the subsequent research, there was no sign for reservation. It was widely believed that this increase in the joint probability is always valid, and this is the source of bosons bunching.

This conclusion is indeed valid provided both amplitudes have no zero points. In appendix A we prove that at every spatial location where both amplitudes do not vanish, then Feynman's reasoning is valid for a point detector. However, if, only *one* of the amplitudes, say $a$, vanishes at a specific point in space, and the two detectors are located on its either sides, then $a_1 \cong -a_2 \cong a$, while $b_1 \cong b_2 \cong b$ (we elaborate on that in the following sections), and

$$p = |a_1 b_2 + a_2 b_1|^2 \cong |ab - ab|^2 = 0, \tag{5}$$

which means that despite the fact that the particles are identical bosons, the probability for these nearby detectors to detect them simultaneously can be arbitrarily small. This is an anti-bunching conduct, which contradicts Feynman assertion. In these scenarios bosons exhibit repletion conduct.

It should be noted that this argument is valid no matter how small or narrow the detectors are, provided they are located on two sides of the zero point.

Moreover, it will be shown below that not only that the probability for simultaneous boson detection vanishes at certain scenarios, but that it vanishes faster than the probability to detect simultaneously non-identical particles at the same scenarios.

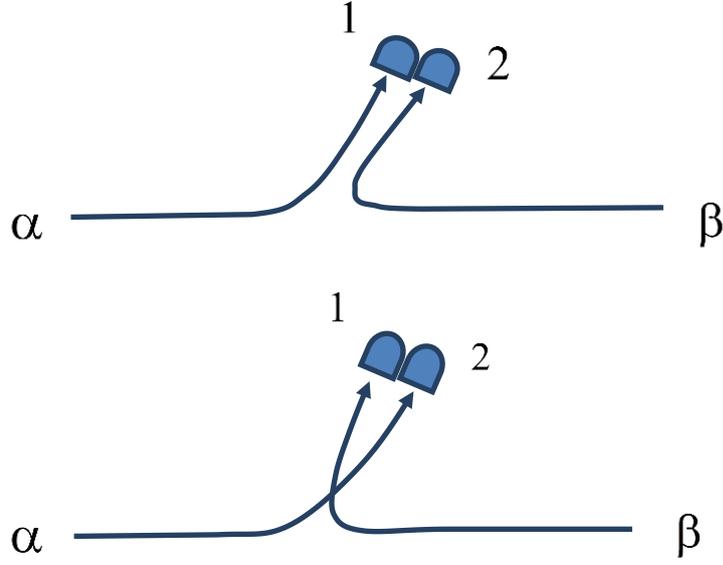

Figure.1: Two scenarios for double scattering into nearby detectors

**Scattering amplitude which zero point – The two point detector case**

Let the sources have profile amplitudes $\psi_1(x)$ and $\psi_2(x)$, i.e., the one-particle probability to find a particle in an infinitesimally small region $dx$ around $x$ is $|\psi_1(x)|^2 dx$ for the first source and $|\psi_2(x)|^2 dx$ for the second source.

Now, to measure joint probabilities, let us place at any given point in space $x$, two detectors at $x \pm \varepsilon$, where it is assumed that $\varepsilon$ is considerably smaller than the characteristic length scale of changes in the wave-functions $\psi_1$ and $\psi_2$ (this is equivalent to Feynman's assumption).

Hence, the joint probability to detect the two particles in case of non-identical (NI) particles is

$$P_{NI}(x)dx = \left\{|\psi_1(x-\varepsilon)\psi_2(x+\varepsilon)|^2 + |\psi_2(x-\varepsilon)\psi_1(x+\varepsilon)|^2\right\}dx \tag{8}$$

and if the particles are identical (I) bosons, is

$$P_I(x)dx = |\psi_1(x-\varepsilon)\psi_2(x+\varepsilon) + \psi_2(x-\varepsilon)\psi_1(x+\varepsilon)|^2 dx. \tag{9}$$

The ratio $\rho_B$ between them (the subscript B stands for the case of bosons)

$$\rho_B(x) \equiv \frac{P_I(x)dx}{P_{NI}(x)dx} = \frac{|\psi_1(x-\varepsilon)\psi_2(x+\varepsilon) + \psi_2(x-\varepsilon)\psi_1(x+\varepsilon)|^2}{\{|\psi_1(x-\varepsilon)\psi_2(x+\varepsilon)|^2 + |\psi_2(x-\varepsilon)\psi_1(x+\varepsilon)|^2\}} \qquad (10)$$

is not a constant 2.

In fact, let us further assume that at $x = x_0$, one, and only one, of the amplitudes vanishes, say $\psi_1(x_0) = 0$. That is, without loss of generality

$$\psi_2(x \cong x_0) = C \qquad (6)$$

and

$$\psi_1(x \cong x_0) = c(x - x_0), \qquad (7)$$

where $C$ and $c$ are complex constants. We will discuss high order zeros, such as $\psi_1(x \cong x_0) = c(x - x_0)^n$ in appendix B.

At the vicinity of $x \cong x_0$ the ratio converges to a generic function, which is independent of the specific features of the wave-functions (on the first approximation)

$$\rho_B(x) = 2\left[1 - \frac{1}{(x-x_0)^2/\varepsilon^2 + 1}\right]. \qquad (8)$$

This Lorenzian function is plotted in Fig.2 (solid line). Obviously, it is accurate only at the vicinity of $x_0$, but it is a good approximation even for $|x_0 - x| \gg \varepsilon$, since the function converges quickly (with the scale of $\varepsilon$) to the value 2, provided $x$ is not close to another zero point.

In case the mean distance between adjacent zero points is $\Delta x$, then one can define an average $\rho_B$, which is approximately

$$\bar{\rho}_B = \frac{1}{\Delta x}\int_x^{x+\Delta x}\rho_B(y)dy \cong 2\left(1 - \pi\frac{\varepsilon}{\Delta x}\right) \qquad (9)$$

which is always lower than 2.

It can easily be shown, that in the case of fermions, where $P_I(x)dx = |\psi_1(x-\varepsilon)\psi_2(x+\varepsilon) - \psi_2(x-\varepsilon)\psi_1(x+\varepsilon)|^2 dx$ the ratio is

$$\rho_F(x) \equiv \frac{P_I(x)dx}{P_{NI}(x)dx} = \frac{|\psi_1(x-\varepsilon)\psi_2(x+\varepsilon) - \psi_2(x-\varepsilon)\psi_1(x+\varepsilon)|^2}{\{|\psi_1(x-\varepsilon)\psi_2(x+\varepsilon)|^2 + |\psi_2(x-\varepsilon)\psi_1(x+\varepsilon)|^2\}}, \quad (10)$$

which, at the vicinity of the zero point satisfies

$$\rho_F(x) = 2 - \rho_B(x) = \frac{2}{(x-x_0)^2/\varepsilon^2 + 1}, \quad (11)$$

and its average value is

$$\bar{\rho}_F = \frac{1}{\Delta x} \int_x^{x+\Delta x} \rho_F(y) dy \cong 2\pi \frac{\varepsilon}{\Delta x}. \quad (12)$$

Clearly, $\frac{1}{2}[\rho_F(x) + \rho_B(x)] = 1$ everywhere.

However, we see that at the vicinity of the zero point they exchange conduct: fermions behave like bosons (bunching) and bosons like fermions (anti-bunching).

**Scattering amplitude with zero point – The finite width detector case**

Feynman also showed that his explanation for bosons bunching applies for a finite width detector. That is, we replace the two point detectors with a single two-particle detector. Such a physical detector can be based on two photons microscopy [14]

In this case, we still assume that $\psi_2(x \cong x_0) = C$ and $\psi_1(x \cong x_0) = c(x-x_0)$ but we now have two variables $z$ and $y$, which have to be integrated over the detector width $[-\varepsilon, \varepsilon]$. The ratio in this case is

$$\rho_B(x) = \frac{\int_{x-\varepsilon}^{x+\varepsilon} [(z-x_0) + (y-x_0)]^2 dzdy}{\int_{x-\varepsilon}^{x+\varepsilon} [(z-x_0)^2 + (y-x_0)^2] dzdy}, \quad (13)$$

which is equal to

$$\rho_B(x) = \frac{1 + 6(x-x_0)^2/\varepsilon^2}{1 + 3(x-x_0)^2/\varepsilon^2} \tag{14}$$

This ratio is always larger than 1, and at $|x - x_0| \gg \varepsilon$ converging to the bunching ratio 2, but at the vicinity of the zero point, it has the value 1 *as if the particles are non-identical* (see dashed line on Fig.2).

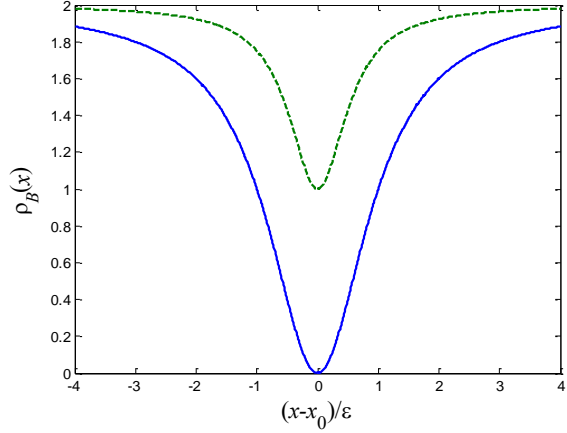

Figure 2. Comparison between the ratio $\rho_B(x)$ in the two point-detector case (solid line), and the wide two-particle detector (dashed line).

## Practical example: Two-sources experiments

This effect can be found in many experiments. In this section we present two scattering experiments. In both of them there are two sources and two detectors (Fig.3), however in the first the effect (bosons anti-bunching) is missing, while in the second it appears.

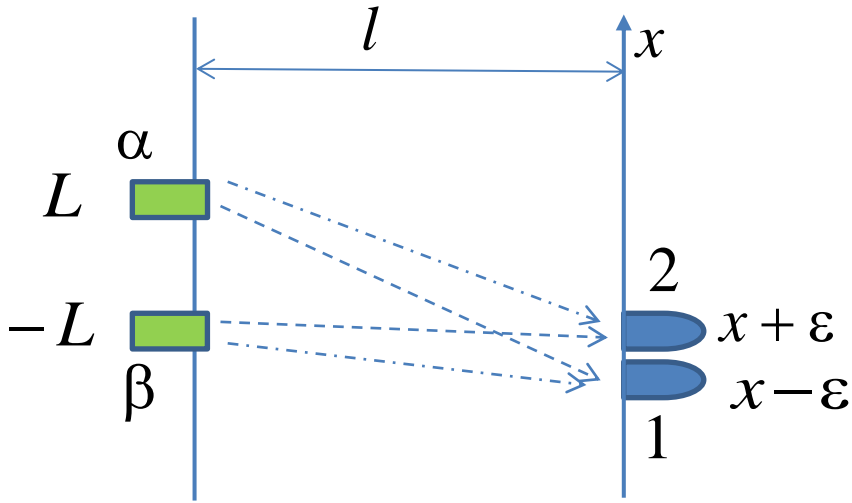

Figure 3: Two source experiment: System schematic

In the first example we choose that the sources have Gaussian profiles, and therefore their far field (on the measuring screen –see Fig.3) wave-functions are also Gaussian

$$\psi_1(x) = (\xi\pi)^{-1/4} \exp(-(x-L)^2/2\xi^2) \tag{14}$$

and

$$\psi_2(x) = (\xi\pi)^{-1/4} \exp(-(x+L)^2/2\xi^2), \tag{15}$$

i.e. the distance between the two sources is $2L$ and the FWHM of the single-particle probability on the screen is $2\xi\sqrt{\ln 2}$.

It should be noted that $\psi_1(x)$ and $\psi_2(x)$ are the wave-functions *on the screen*, which, in the case where the distance to the screen is extremely large (i.e. $l \gg |x|, L, \xi$), means that the sources are also Gaussian, with $\psi_{source}(x) \propto \exp(-(x-L)^2/2\sigma^2)$, where $\sigma \cong l/k\xi = \lambda l/2\pi\xi$ ($k$ and $\lambda$ are approximately the mean wave number and wavelength of the particles' beam).

In this case, since both wave-functions have no zeros, then $\rho \cong 2$, and in fact $\lim_{\varepsilon \to 0} \rho = 2$ (see Fig.4 and the upper panel of Fig.6).

However, when the sources profile have a rectangular shape, then their far field wave-functions density has a "sinc" shape

$$\psi_1(x) = \sin[\pi(x-L)/\xi]/[\pi(x-L)/\xi] \tag{16}$$

and

$$\psi_2(x) = \sin[\pi(x+L)/\xi]/[\pi(x+L)/\xi]. \tag{17}$$

Again, in the case where the distance to the screen is extremely large (i.e. $l \gg |x|, L, \xi$), this means that the source have approximately a rectangular shape with width $\lambda l/\xi$, i.e., $\psi_{source}(x) \propto \text{rect}_{\lambda l/\xi}(x)$.

In this case both wave-functions have an infinite number of zeros at $x_m^{(1)} = m\xi + L$ for $\psi_1(x)$ and $x_m^{(2)} = m\xi - L$ for $\psi_2(x)$ (where in both cases $m$ is a non-zero integer).

Since in most cases (unless $\tilde{m}\xi = L$, for any integer $\tilde{m}$), there are infinite points for which $\psi_1(x) = 0$ but $\psi_2(x) \neq 0$ or vice versa $\psi_2(x) = 0$ but $\psi_1(x) \neq 0$. At the vicinity of all these points we will find the conduct presented above, where $\rho < 2$ (see Fig.5 and the lower panel of Fig. 6). Moreover, since the mean distance between adjacent zeros is $\xi/2$ (since both functions have zeros) then the mean value of $\rho$ is

$$\bar{\rho} \cong 2\left(1 - \pi\frac{2\varepsilon}{\xi}\right). \tag{18}$$

In Fig.7 a similar comparison is presented for fermions. Clearly, at the points where bosons experience anti-bunching, fermions bunch and vice-versa.

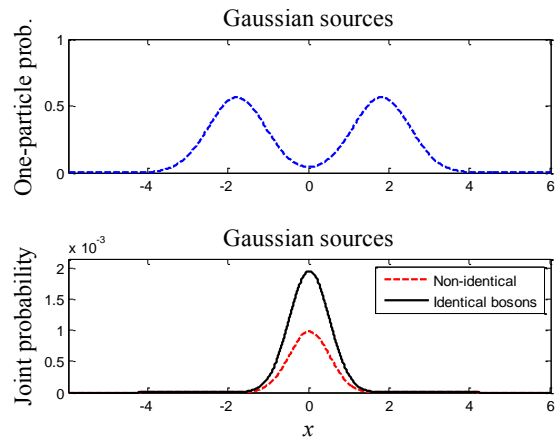

Figure 4: The one-particle probability density (upper panel) and the joint probability density (lower panel) when the sources have Gaussian profiles.

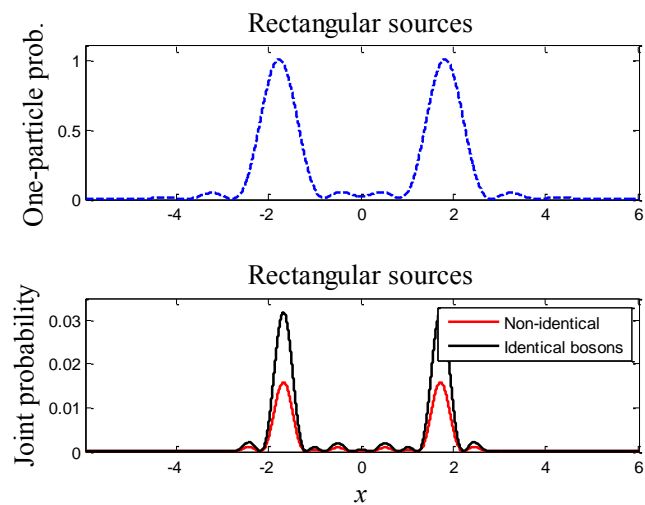

Figure 5: The one-particle probability density (upper panel) and the joint probability density (lower panel) when the sources have rectangular profiles.

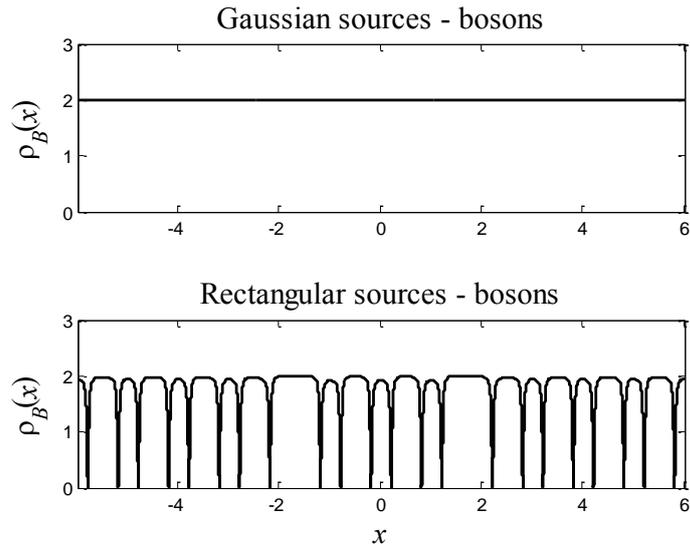

Figure. 6: The ratio between the joint probability of non-identical particles and the joint probability of identical bosons. In the upper panel the sources have a Gaussian profile, while in the lower panel they have a rectangular profile.

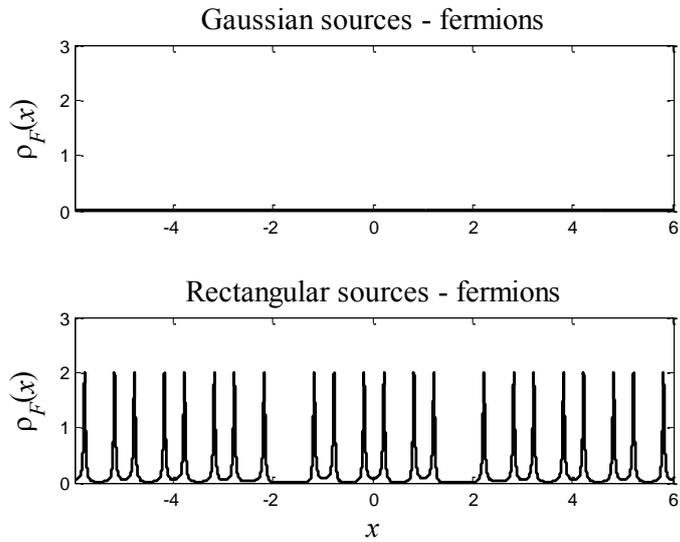

Figure 7: The same as Fig.6 but for fermions.

## Experimental Atomic Bunching and Anti-Bunching in the Literature

It is well known that boson bunching (and fermion anti-bunching) was revealed in numerous experiments as the atomic analog of the HBT experiment. For example, in Refs.[15 and 16], a falling cloud of parahelium atoms exhibited a clear bunching conduct.

In these experiments the initial state of the cloud was approximately Gaussian (due to the specific initial harmonic trapping). In order to measure localized anti-bunching (as is presented in this paper) the researcher should consider different trapping shapes, like rectangular traps or passing the cloud through double-two-slits (i.e., passing the left side of the cloud through two-slits, while passing the its right side through another two-slits). In any case, since the anti-bunching appears only at the vicinity of the places where one of the wave-functions vanishes, averaging would kill the effect. The measurement should take place only at the zeros proximity.

## Discussion and Summary

The argument given by Feynman for bosons bunching is revisited. In his original wording: "It is a property of Bose particles that if there is already one particle in a condition of some kind, the *probability* of getting a second one in the same condition is twice as great as it would be if the first one were not already there"[1] (see also the quotation in the abstract, and Ref.[4]).

By a spatial analysis of the scenario suggested in his well-known lectures, it is shown that Feynman's argument is only partially correct, and as a general rule does not hold.

Moreover, it can easily be demonstrated that in certain cases the exact opposite occurs: the probability that a couple of bosons would arrive at a certain place can be lower than the probability that two non-identical particles would arrive there. This behavior can be regarded as an anti-bunching conduct of bosons. The source of this special conduct lies in the presence of zeros in the SPWF.

If one of the SPWF vanishes at certain spatial point then the bosons cannot bunch in its vicinity. If the product of both SPWF changes sign with respect to the zero point, then the bosons will behave there like non-identical particles, but if the product changes sign then they will anti-bunch like fermions.

From similar reasoning, at the vicinity of zeros in the SPWF fermions will bunch like bosons.

A simple scattering experiment is suggested to validate this effect.

**Appendix A:**

In this appendix we show that $\lim_{\varepsilon \to 0} \rho_B(x) < 2$ (or, if the distance between the detector is always finite then $2 - \rho_B(x) > O(\varepsilon^2)$), only if either $\psi_1(x_0) = 0$ or $\psi_2(x_0) = 0$ at a certain point $x_0$.

Let us assume that this condition does not hold, i.e., $\psi(x_0) \neq 0$ and $\psi(x_0) \neq 0$. Let us further define a new function

$$f(x) \equiv \psi_1(x_0 - x)\psi_2(x_0 + x) \tag{A1}$$

According to the basic assumption $f(x) \neq 0$ as well then,

$$\rho_B(x_0) \equiv \frac{|\psi_1(x_0 - \varepsilon)\psi_2(x_0 + \varepsilon) + \psi_2(x_0 - \varepsilon)\psi_1(x_0 + \varepsilon)|^2}{\{|\psi_1(x_0 - \varepsilon)\psi_2(x_0 + \varepsilon)|^2 + |\psi_2(x_0 - \varepsilon)\psi_1(x_0 + \varepsilon)|^2\}} =$$

$$\frac{|f(\varepsilon) + f(-\varepsilon)|^2}{|f(\varepsilon)|^2 + |f(-\varepsilon)|^2} = 2 - \frac{|f(\varepsilon) - f(-\varepsilon)|^2}{|f(\varepsilon)|^2 + |f(-\varepsilon)|^2} \tag{A2}$$

$$\rho_B(x_0) = 2 - \frac{|f(\varepsilon) - f(-\varepsilon)|^2}{|f(\varepsilon)|^2 + |f(-\varepsilon)|^2} \cong 2 - \varepsilon^2 \frac{|f'(0)|^2}{2|f(0)|^2} \tag{A3}$$

and hence

$$\rho_F(x_0) = 2 - \rho_B(x_0) \cong \varepsilon^2 \frac{|f'(0)|^2}{2|f(0)|^2} \tag{A4}$$

Q.E.D.

**Appendix B: Higher order zeros**

In general, the wave-function can vanish like

$$\psi_1(x \cong x_0) = c(x - x_0)^n, \tag{B1}$$

for $n > 1$. In this case

$$\rho(x \cong x_0) = \frac{|(x - x_0 - \varepsilon)^n + (x - x_0 + \varepsilon)^n|^2}{|x - x_0 - \varepsilon|^{2n} + |x - x_0 + \varepsilon|^{2n}}, \tag{B2}$$

which exhibit a more complicated structure. Again, this is a generic function (all the dependence on the constants $c$ and $C$ disappeared).

Instead of one length scale ($\varepsilon$), two length scales emerges: $\delta_1 = \varepsilon/(\sqrt{2}n)$ and $\delta_2 = \sqrt{2}n\varepsilon$ (see Fig. B1).

At the vicinity of the zero point $x_0$, $\rho$ varies rapidly (on the length scale of $\delta_1 = \varepsilon/(\sqrt{2}n)$) from 2 (if $n$ is even) or 0 (if $n$ is odd) to 1. Then, $\rho$ increases monotonically on the length scale $\delta_2 = \sqrt{2}n\varepsilon$ to the value 2. That is,

$$\rho(|x - x_0| > \varepsilon) = 2 - \frac{2n/(2n-1)}{(x-x_0)^2/[n(2n-1)\varepsilon^2]+1}, \qquad (B3)$$

and

$$\rho(|x - x_0| < \varepsilon) = \frac{[(-1)^n+1]^2 + n[4n - [(-1)^n+1]^2](x-x_0)^2/\varepsilon^2}{2 + 2n(2n-1)(x-x_0)^2/\varepsilon^2}$$

$$= \begin{cases} 2n^2(x-x_0)^2/\varepsilon^2 & \text{for } n \text{ odd} \\ 2\dfrac{1+n[n-1](x-x_0)^2/\varepsilon^2}{1+n(2n-1)(x-x_0)^2/\varepsilon^2} & \text{for } n \text{ even} \end{cases} \qquad (B4)$$

For large $n$ we can ignore the short scale conduct regime, in which case

$$\bar{\rho} = \frac{1}{\Delta x}\int_x^{x+\Delta x}\rho(y)dy \cong 2 - \pi\frac{\sqrt{2}n\varepsilon}{\Delta x} \qquad (B5)$$

which is larger than $2\left(1 - \pi\dfrac{\varepsilon}{\Delta x}\right)$ for the n=1 case, but still smaller than 2.

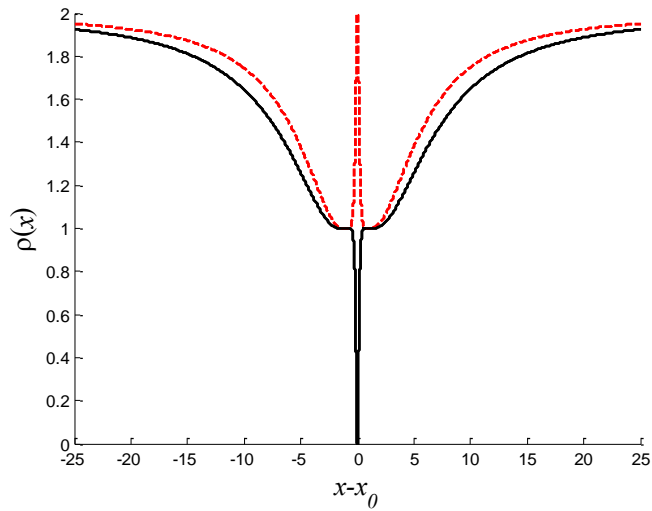

Figure B1: The ratio $\rho(x)$ as a function of the spatial coordinate. The dashed curve stands for *n*=4, and the solid curve for *n*=5.

# References

[1] R.P. Feynman, R. B. Leighton , M. Sands , *The Feynman Lectures on Physics: Quantum mechanics*, Addison-Wesley (1965)

[2] W. Pauli, *Exclusion Principle and Quantum Mechanics*, Nobel Lecture

[3] C. Cohen-Tannoudji, B. Diu, and F. Laloë, *Quantum Mechanics* (John Wiley and Sons, Paris, 1977)

[4] R. Feynman and A. R. Hibbs, *Quantum Mechanics and Path Integrals*, Emended Edition, (Dover Publications inc' Mineola, New York 2005)

[5] U. Fano, "Quantum theory of interference effects in the mixing of light from phase independent sources". American Journal of Physics **29**, 539, (1961).

[6] E. Purcell, Nature **178**, 1449–1450 (1956).

[7] B. L. Morgan and L. Mandel, Phys. Rev. Lett. **16**, 1012–1014 (1966).

[8] R. Hanbury Brown and R. Q. Twiss (1956). "A Test of a New Type of Stellar Interferometer on Sirius". Nature 178 (4541): 1046–1048

[9] G. Alexander, Rep. Prog. Phys. **66,** 481–522 (2003)

[10] T. Jeltes, J. M. McNamara, W. Hogervorst, W. Vassen, V. Krachmalnicoff, M. Schellekens, A. Perrin, H. Chang, D. Boiron, A. Aspect, C. I. Westbrook, **Nature** 445, 402-405  and references therein.

[11] Y. Bromberg, Y. Lahini, E. Small and Y. Silberberg, Nature Photonics **4**, 721–726 (2010).

[12] R. Hanbury Brown and R. Q. Twiss, Proc of the Royal Society of London A **243**, 291–319 (1958).

[13] Hereinafter, for the sake of simplicity, we adopt Feynman's terminology and use the terms: bosons and fermions where it should have been "spinless bosons" and "spinless fermions" respectively or "bosons with symmetric spin function" and "fermions with symmetric spin function" respectively.

[14] W. Denk, J. Strickler, W. Webb. "Two-photon laser scanning fluorescence microscopy", Science **248**, 73–6 (1990)

[15] M. Schellekens, R. Hoppeler, A. Perrin, J. V. Gomes, D. Boiron, A. Aspect, and C. I. Westbrook, Science 310, 648 (2005)

[16] T. Jeltes, J. M. McNamara, W. Hogervorst, W. Vassen, V.Krachmalnicoff, M. Schellekens, A. Perrin, H. Chang, Boiron, A. Aspect, and C. I. Westbrook, ,

"Comparison of the Hanbury Brown-Twiss effect for bosons and fermions," Nature **445**, 402 (2007)